\pdfoutput=1
\documentclass[11pt,a4paper,twoside,twocolumn]{article}
\usepackage[utf8]{inputenc}
\usepackage{amsmath}
\usepackage{amsfonts}
\usepackage{amssymb}
\usepackage{graphicx}
\usepackage{caption}
\usepackage{subcaption}
\usepackage{mathtools}
\usepackage[left=2cm,right=2cm,top=2cm,bottom=2cm]{geometry}

\usepackage[format=plain,justification=raggedright,singlelinecheck=false,font={stretch=1.125,small,sf},labelfont=bf,labelsep=space]{caption}
\usepackage{fnpos}
\usepackage{array}
\usepackage{droidsans}
\usepackage{charter}
\usepackage[T1]{fontenc}
\usepackage{setspace}
\usepackage[compact]{titlesec}

\title{Attraction-induced jamming in the flow of foam through a channel}
\author{Karthik Menon$^a$ , Rama Govindarajan$^a$, and Shubha Tewari$^{*b}$
} 
\date{}

\begin{document}

\twocolumn[
  \begin{@twocolumnfalse}
      \maketitle
      \begin{center}
       $^a$ TIFR Centre for Interdisciplinary Sciences, Tata Institute of Fundamental Research, Narsingi, Hyderabad - 500075, India. \\
$^b$ Physics Department, University of Massachusetts, Amherst, MA 01003, USA. E-mail: tewari@physics.umass.edu
\end{center}           

    \begin{abstract}
      We study the flow of a pressure-driven foam through a straight channel using numerical simulations, and examine the effects of a tuneable attractive potential between bubbles. This potential, which accounts for the effects of disjoining pressure in the liquid films between separating bubbles, is shown here to introduce jamming and stick-slip flow in a straight channel. We report on the behaviour of these new regimes by varying the strength of the attractive potential. It is seen that there is a force threshold below which the flow jams, and on increasing the driving force, a cross over from intermittent (stick-slip) to smooth flow is observed. This threshold force below which the foam jams increases linearly with the strength of the attractive potential. By examining the spectra of energy fluctuations, we show that stick-slip flow is characterized by low frequency rearrangements and strongly local behaviour, whereas steady flow shows a broad spectrum of energy drop events and collective behaviour. Our work suggests that the stick-slip and the jamming regimes occur due to the increased stabilization of contact networks by the attractive potential - as the strength of attraction is increased, bubbles are increasingly trapped within networks, and there is a decrease in the number of contact changes.
    \end{abstract}

  \end{@twocolumnfalse}
]

\section{Introduction}

The flow of liquid foams has been a subject of considerable interest in the recent past due to the multitude of applications in which these materials find use\cite{physfoams}. While much effort has focused on the response of foam to external shear, there are a number of recent studies of pressure or velocity driven foams through channels of different configurations. Of these, many studies on pressure-driven flowing foams explore the effects of channel geometry and properties such as polydispersity on various aspects of the flow. Experiments on the effect of channel width were performed by Jones et al. \cite{jonesphysfluids2013} and Dollet et al. \cite{dolletpre2014}, where they also developed mathematical models for the pressure drop. Flow through channels with a constriction have been studied both experimentally \cite{joneselsevier2011,dolletjrheol2010} and numerically \cite{langloisjrheol2014}. The flow of foam around a circular obstacle has also been an area of study by the group of Graner \cite{cheddadiepje2011,dolletjfm2007}, who have developed constitutive relations for bubble deformations based on experiments and simulations. A more recent experimental and simulational study \cite{dolletJFM2015} of plastic rearrangement events in the channel flow of a foam shows that wall friction plays a role in the shape of the velocity profiles, and indicates a link between these local rearrangements and the macroscopic rheology.

In this paper, we examine the pressure-driven flow of foam in a parallel channel. Rather than focus on the effects of geometry, as has been done previously, we simulate a more realistic model foam which produces a new flow regime even in this simple system. The new ingredient in our model, which is based on Durian's bubble model\cite{durianPRL1995}, is an attractive interaction that takes into account the effects of disjoining pressure, and we examine the effects of tuning the strength of this attraction relative to the repulsion. We find that the presence of an attractive force extends the range of driving forces over which jamming occurs, and introduces a new stick-slip regime, from whence the flow makes a transition to steady flow as the driving is increased. The threshold driving force for unjamming increases linearly with the strength of the attraction. In the stick-slip regime, strong localized rearrangements occur after stress builds up over long time scales, and the kinetic and stored elastic energy of attraction are of the same order of magnitude. Conversely, there are constant rearrangements in the steady flow regime and the kinetic energy dominate over the attraction.

There has been previous work on the effect of an attractive potential on the flow and jamming of sheared foams, but there has been little interest in pressure-driven channel flow, with some exceptions discussed below. Theoretical results from Denkov et al \cite{denkovprl2009} show that the spontaneous thinning of films, which gives rise to an adhesion between bubbles, leads to the jamming of sheared foams. Chaudhuri et al \cite{PinakiPRE2012,PinakiPRL2014} showed that the inclusion of an attractive potential in a model for soft glassy systems leads to inhomogeneities in sheared foams, particularly close to the yield limit. In the context of channel flow, there has not been much focus on the effect of the attraction. Chaudhuri and Horbach \cite{PinakiPRE2014Channel} have looked at Poiseuille flow, using a model for confined soft glasses with Yukawa-like interactions, and demonstrated the relation of the width of the flow geometry to stress inhomogeneity and the avalanche-like onset of flow in the channel. Their work did not address the role of an attractive potential in the flow dynamics.  When modeling the pressure-driven foam flow through a narrow constriction using the original Durian model\cite{durianPRL1995}, without attraction, Langlois \cite{langloisjrheol2014} suggested that the discrepancy between simulation and experiment seen in the expanding flow of foam might be remedied by including the effect of disjoining pressure. But there has not yet been an implementation of that suggestion. Here, our primary emphasis is on how the strength of the attraction affects the onset of flow and the subsequent response of the system to external driving in parallel channel flow.

Foams are thought to be Herschel Bulkley fluids, exhibiting a finite yield stress and a non-linear stress-strainrate behaviour.  Experimental \cite{gopaljcolloid1999,ovarlez2008wide} and theoretical \cite{denkovprl2008,tcholakovapre2008} studies of foam under steady shear have elucidated the coupling between film-level dynamics and the macroscopic shear response. Understanding this coupling has provided insight into the scaling laws for viscous dissipation and wall-shear in flow through narrow channels \cite{dolletjfm2010,terriacepl2006,cantatepl2004,raufastephysfluids2009}, though these studies have focused primarily on the steadily flowing regime. The jamming behaviour of foams has received a fair bit of attention too, but almost exclusively in the shear-driven scenario \cite{tighe2011jamming,tighe2010jamming,van2014contact}. Examining the changing response of foam as the gas fraction is varied from the wet foam limit to the finite yield stress state, Katgert et al. \cite{katgert2013} point out that a signature of the non-trivial rheology of foam is an increase in velocity fluctuations with decreasing flow rate close to the jamming transition, and propose scaling laws for the flow close to jamming. There has been, however, minimal attention given to jamming in channel flow of foams, which is an important geometry to consider for flowing applications. Understanding this is hence our primary motivation behind this work. 

To explore a regime of response ranging from jammed to stick slip to smooth flow, we use a modified version of the Bubble Model \cite{durianPRL1995}, a bubble-scale numerical model which has been shown to work well for wet foams in the flowing regime \cite{durianpre1997,tewaripre1999,langloispre2008}. Many numerical models previously developed for the study of foams at or near the dry limit are difficult to extend beyond the quasistatic regime. In this regime, foam has been modeled as a network of vertices, plateau borders, and films \cite{weairephil1983,weairephil1984}, an approach that works best at or near the dry limit. The effects of increasing liquid fraction have been accounted for by including viscous dissipation as in the vertex model \cite{okuzonojrheol1993}, or curvature and liquid content in the plateau borders \cite{kernpre2004} using the Surface Evolver software \cite{brakke1992} to model the evolution of the foam at each step. More recently, lattice models such as the lattice Boltzmann method \cite{dolletJFM2015} allow an exploration of the dynamics of a driven foam while incorporating dissipative mechanisms and boundary effects. Most models that allow for dynamical studies have ignored the effects of the attractive force between films caused by disjoining pressure, which becomes especially relevant at low driving forces in channel flow. With our inclusion of the attraction potential in a model that is not restricted to quasistatic dynamics, we are able to explore different regimes of foam flow that previous studies have not addressed. 

We next describe our modified model foam and the details of the numerical simulation, and discuss our results and conclusions in the following sections.

%\begin{figure*}
% \centering
% \def\svgwidth{1.3\columnwidth}
% \input{schematics.pdf_tex}
% \caption{(Left) Schematics showing the shear and overlap in the bubble model. (Right top) A schematic of the computational domain used in this work. The driven portion of the channel is of length $H$. The three $x$ positions marked refer to the 3 locations at which the mean velocity in Figure \ref{bejan_vs_meanvel} are shown, and correspond to the $x$ coordinates measured from the end of the driven portion scaled by the length $(L_x - H)$. (Right bottom) A representative snapshot of the simulation at the initial configuration. The bubbles highlighted in red are held in place to create walls and have radii equal to $\langle R \rangle$.}
% \label{schematics}
%\end{figure*}

\begin{figure*}
\includegraphics[scale=.35]{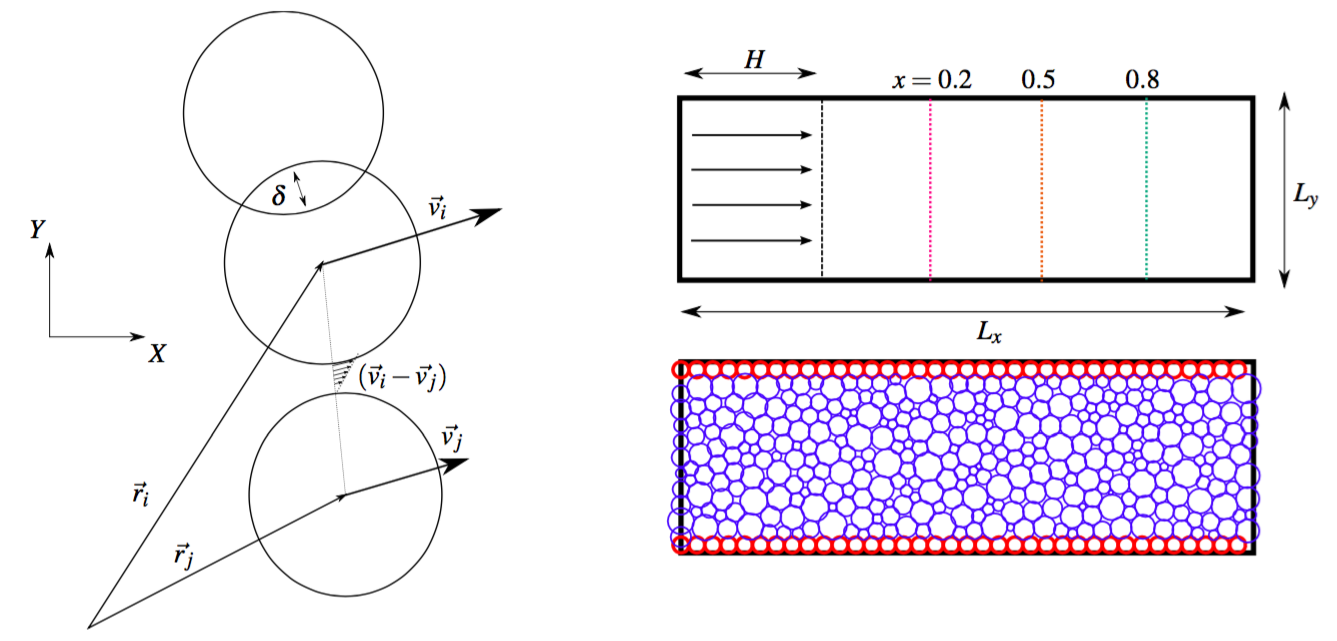}
\caption{(Left) Schematics showing the shear and overlap in the bubble model. (Right top) A schematic of the computational domain used in this work. The driven portion of the channel is of length $H$. The three $x$ positions marked refer to the 3 locations at which the mean velocity in Figure \ref{bejan_vs_meanvel} are shown, and correspond to the $x$ coordinates measured from the end of the driven portion scaled by the length $(L_x - H)$. (Right bottom) A representative snapshot of the simulation at the initial configuration. The bubbles highlighted in red are held in place to create walls and have radii equal to $\langle R \rangle$.}
\label{schematics}
\end{figure*}

\section{Numerical Description}
\subsection{The Bubble Model}

We use a modified version of Durian's Bubble Model \cite{durianpre1997}, in which the detailed geometry of liquid films and Laplace borders is replaced by a collection of bubbles, or soft particles, with pairwise interactions. This model is particularly useful for studying the dynamics of a disordered wet foam in response to external driving. In two dimensions, bubbles are modelled as soft circular disks, with overlapping boundaries representing bubble deformation. The overlapping bubbles experience a harmonic repulsion, a valid approximation to the effective interaction due to deformation in two dimensions; each bubble also experiences a viscous force proportional to the velocity difference between it and its neighbours. The channel is taken to be horizontal, so there are no drainage effects in the liquid films. We also ignore the effects of diffusive coarsening and assume there is no coalescence of bubbles; this is reasonable as diffusion occurs on a much longer timescale than the typical timsescale associated with the driving.

In the original model, the repulsive force on bubble $i$, due to each overlapping neighbour, $j$, is given by a constant of proportionality times the degree of overlap $\delta_{ij}$: 
\begin{equation}
\vec{F_{ij}^r} = k_r \delta_{ij} \frac{(\vec{r_i}-\vec{r_j})}{|\vec{r_i}-\vec{r_j}|},
\end{equation}
with 
\begin{equation}
\delta_{ij} = (R_i + R_j) - |\vec{r_i} - \vec{r_j}|.
\end{equation}
The effective spring constant $k_r$ depends on Laplace pressure differences between the bubbles, and the strength of the repulsion, $F_0^r$, is a constant that depends on the surface tension between the liquid-gas phases in the foam, and is determined in part by the surfactant properties and concentration:
\begin{equation}
k_r = \frac{F_0^r}{(R_i + R_j)}.
\end{equation}

We now describe the two modifications we have made here to enhance the model's validity. The first is to extend the range of the viscous interaction. Durian's original model has viscous forces only between overlapping bubbles, which are proportional to the relative velocity:
\begin{equation}
\vec{F_{ij}^v} = -b(\vec{v_i}-\vec{v_j}).
\end{equation}
This assumes a constant film size that scales as ${\langle R \rangle}^2$ for a bubble of radius $R$, so that the constant of proportionality, $b$, depends only on the viscosity of the intervening fluid.

Our modification extends the viscous interaction to non-contacting nearest neighbors, within a small region around a bubble. This accounts for the effects of shear within the intervening fluid. This force drops off inversely with the separation between the bubble centres. Hence the viscous force is reframed as
\begin{equation}
\vec{F_{ij}^v} =
\begin{dcases*}
-b(\vec{v_i}-\vec{v}_j)\frac{\eta}{\eta-\delta_{ij}} &
  for $\delta_{ij} \leq 0$;\\
-b(\vec{v_i}-\vec{v_j})& for $\delta_{ij} > 0$;
\end{dcases*}
\end{equation}
where $\eta$ is a small, non-zero parameter.

The second, more significant addition to the model, is a short-ranged harmonic attraction.  This accounts for the effect of a positive disjoining pressure, which can have a substantial effect on the dynamics of real foams. As neighbouring bubbles lose contact, the spring force becomes attractive, but drops to zero when the bubble separation exceeds $\epsilon$, set to a tenth of the mean bubble radius. The attractive force is written as
\begin{equation}
\vec{F_{ij}^a} =
\begin{dcases*}
k_a \delta_{ij} \frac{(\vec{r_i}-\vec{r_j})}{|\vec{r_i}-\vec{r_j}|} &
  for $\delta_{ij} \leq \epsilon$;\\
0 & for $\delta_{ij} > \epsilon$;
\end{dcases*}
\end{equation}
where the attractive harmonic constant, set to $k_a = \alpha k_r$, is varied relative to the repulsive spring constant $k_r$, with $\alpha$ a adjustable parameter in this study. 

We now calculate the behaviour of every bubble by a vector sum of the forces acting on the bubble. Foams are highly overdamped, and the net force on each bubble should be zero. For computational simplicity, though, we assign a small mass to each bubble, which allows the use of time-marching techniques to integrate the equations of motion rather than the inversion of a sparse matrix at each step.  The equation of motion of each bubble is given by:
\begin{equation}
m\cdot\frac{d\vec{v_i}}{dt} = \sum_{j}(\vec{F_{ij}^r} + \vec{F_{ij}^v} + \vec{F_{ij}^a})+\vec{F}_{i}^{p}
\label{eqn_motion}
\end{equation}
where $F_{i}^{p}$ is an externally applied force which we describe in the next section. Thus given the position vectors and radii of each bubble at any instant of time, this model provides a simple way to study the behaviour of the foam at subsequent times. 

\subsection{Computational Details}

Our computational domain consists of a straight channel, as shown in Figure \ref{schematics}, of width $50\langle R \rangle$ and length $150\langle R \rangle$, where $\langle R \rangle$ is the mean radius of the bubbles in the system. The walls are lined with a row of bubbles of radius $\langle R \rangle$, that are fixed in place during the entire duration of the simulation. Thus the interaction between the boundary walls and the bulk are determined by bubble-bubble interactions. Bubble radii are chosen from a triangular distribution which is centered at $\langle R \rangle$ and whose width is determined by the polydispersity, $w$, which we keep fixed at a constant value $w = 0.75$. The mean radius $\langle R \rangle$ is calculated by fixing the gas fraction, $\phi = 0.95$ and the total number of bubbles in the domain, $N_x\times N_y = 1000$. Thus, 
\begin{equation}
\phi = \frac{(N_xN_y)\cdot\langle A \rangle}{(L_xL_y)} = \frac{\pi(N_xN_y)\langle R\rangle^2[1+w^2/6]}{(L_xL_y)}.
\end{equation}

After initializing the bubbles on a triangular lattice, the foam is equilibrated by iterating equation \ref{eqn_motion}, without the viscous term as suggested by earlier users of this model, over all bubbles initially placed on the triangular lattice. This is carried out until all bubbles are at equilibrium, i.e. their displacements drop below a threshold, which we set at $\langle R \rangle \times 10^{-10}$. In our simulation, equation \ref{eqn_motion} is integrated using a Crank-Nicholson semi-implicit discretization scheme: 
\begin{equation}
\vec{r_i}(t+dt) = \vec{r_i}(t) + \frac{1}{2}[\vec{v_i}(t) + \vec{v_i}(t+dt)],
\end{equation}
where $\vec{v_i}(t)$ and $\vec{v_i}(t+dt)$ are calculated using a Predictor-Corrector method. The reasoning behind the use of this numerical technique is because it is symmetric over the duration of each time step, and is more stable than the conventional Euler scheme. The integration time step is set to one percent of the characteristic relaxation time of the foam, $\tau$ arising from a balance between elastic and viscous forces:
\begin{equation}
\tau = b\langle R \rangle / F_0.
\end{equation}

The method used to drive the foam mimics an experimental method \cite{CantatPRE2006} to create flowing foams, and is numerically implemented in a similar manner to Langlois \cite{langloisjrheol2014}. By this method, bubbles within a fixed length, $H$, in the initial part of the channel are driven by a force $F^p$. The bubbles in the rest of the channel do not experience this driving force and the flow is brought about by the influence of these driven bubbles on the remaining ones. This generates an effective pressure gradient across the channel, which is the driving mechanism for the flow. By this method, the driving force per unit width across the channel is $F^p \cdot N_d/L_y$, where $N_d$ is the number of bubbles in the driven region. This is expressed in dimensionless form, scaled by the characteristic length scale $\langle R \rangle$ and force $F_0$, as follows:
\begin{equation}
F_d = \frac{(F^p \cdot N_d)/L_y}{F_0/ \langle R \rangle}.
\end{equation}    

The boundary conditions on the inlet and outlet of the channel are periodic. This is enforced to conserve the set of bubbles drawn from the distribution, and hence to preserve the gas fraction.

\section{Results and Discussion}

\subsection{Steady Flow}

Of primary interest for any kind of flowing medium is the velocity and flux behaviour it exhibits. We first present these, and this gives some insight into how our attractive potential manifests itself. 

\begin{figure}
\centering
	\includegraphics[scale = 0.4]{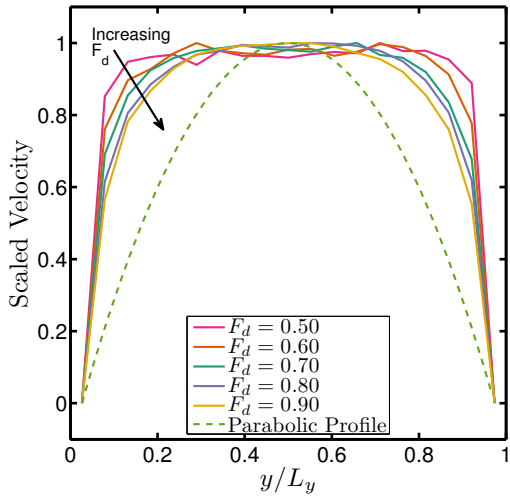}
	\caption{Velocity profiles, scaled by the centerline velocity, show the evolution, with increase in forcing $F_d$, from a plug flow to a shear flow, with the shear permeating farther away from the walls. The parabolic profile corresponds to Poiseuille flow of a Newtonian fluid.}
	\label{velpro_vs_drive_scaled}
\end{figure}

\begin{figure}
\centering
\includegraphics[scale = 0.40]{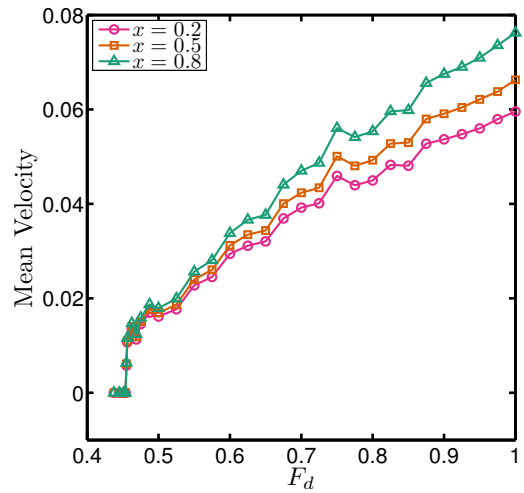}
\caption { Mean velocity as a function of the forcing at three different positions along the centerline of the channel.  It can be seen that there is a downstream increase in centreline velocity, which is more pronounced for higher values of $F_d$. Here, $F_0^a = F_0^r$.}
\label{bejan_vs_meanvel}
\end{figure}

Figure \ref{velpro_vs_drive_scaled} shows time-averaged velocity profiles across the channel during steady flow, scaled by the value at the center of the channel. The velocity profiles resemble those expected for a Bingham plastic fluid, but do not scale at different driving forces. The width of the plug flow region decreases with increasing driving force, as the effect of shear at the walls permeates towards the centreline of the flow. It is clear that the velocity profiles in this range of driving are far from the Poiseuille profile of a Newtonian fluid. These shapes are well described by the exponential fitting function of experimental velocity profiles in Poiseuille flow of foams of Dollet et al. \cite{dolletJFM2015}

To understand the spatial nature of the flow in the channel, Figure \ref{bejan_vs_meanvel} shows the average centreline velocity at three points along the channel. The fact that the average centre-line velocity is the lowest close to the driven region, and highest at the far end of the channel seems counterintuitive at first. This behaviour can be attributed to the denser packing and deformation of bubbles in the region immediately downstream of the driven region. We have verified that the mass flux remains constant through the channel. As expected, the downstream increase in centre-line velocity is smaller at low driving forces since the compression close to the driven region is less pronounced. In order to understand this behaviour, movies of the flow and contact force networks (Electronic Supplementary Information) were studied and it is seen that the region closest to the driven region is a region of densely packed contact networks. Hence, due to the fact that the attractive force serves to render these contact networks more persistant, they form a region of dense obstructions to the flow close to the driven region, while the flow further downstream faces less resistance owing to sparser contact networks.

\subsection{The approach to jamming}

We now describe the flow response in the presence of the attractive potential, and in particular, its role in the dynamics at low driving forces. To demonstrate this, we vary the harmonic constant, $F_0^a$, relative to the strength of the repulsion, and gauge its effect on the flow. The original model, as well as present simulations, show a yield stress below which there is no flow. This is consistent with observations in real foams. Beyond this yield point, our simulations indicate that the flow is steady in the absence of an attractive potential, whereas real foams are expected to have an intermediate regime of stick-slip flow,  dominated by intermittent avalanche-like events. In a flow through a constriction, Langlois \cite{langloisjrheol2014} obtained intermittent flow numerically without an attractive potential, but that was a consequence of the flow geometry. Below we show that an attraction between bubbles gives rise to new intermittent flow regimes even in a straight channel geometry. We also explore the effect of varying the attraction on the behaviour in these regimes.

\begin{figure}[h]
\centering
\includegraphics[scale=0.4]{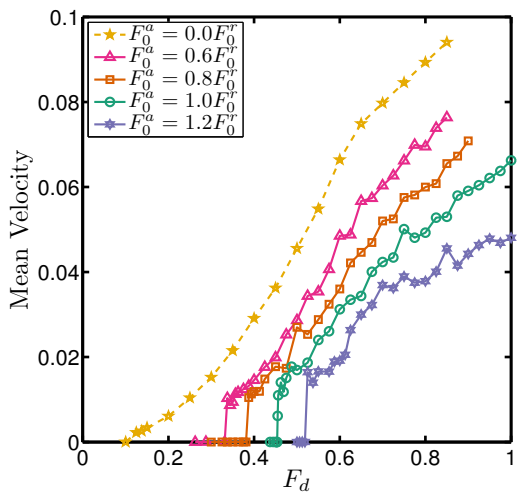}
\caption{ Centerline velocity versus $F_d$ at $x = 0.5$, plotted for different values of the attractive constant, $F_0^a$. The dashed line corresponds to the Bubble Model with zero attraction.}
\label{meanvel_vs_attr}
\end{figure}

\begin{figure}[h]
\centering
\includegraphics[scale=0.4]{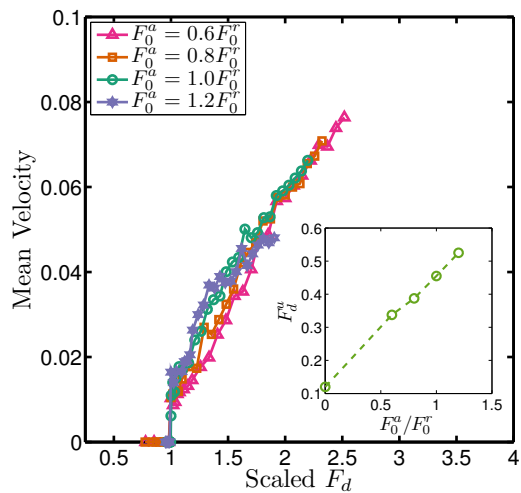}
\caption{ The mean velocity from Figure \ref{meanvel_vs_attr}, as a function of $F_d/F_d^u$, which is the driving force scaled by the corresponding unjamming driving force for each value of attraction. The inset shows the unjamming force to be linearly related to the attractive force. }
\label{unjam_vs_attr}
\end{figure}

\begin{figure}[h]
\centering
\includegraphics[scale=0.43]{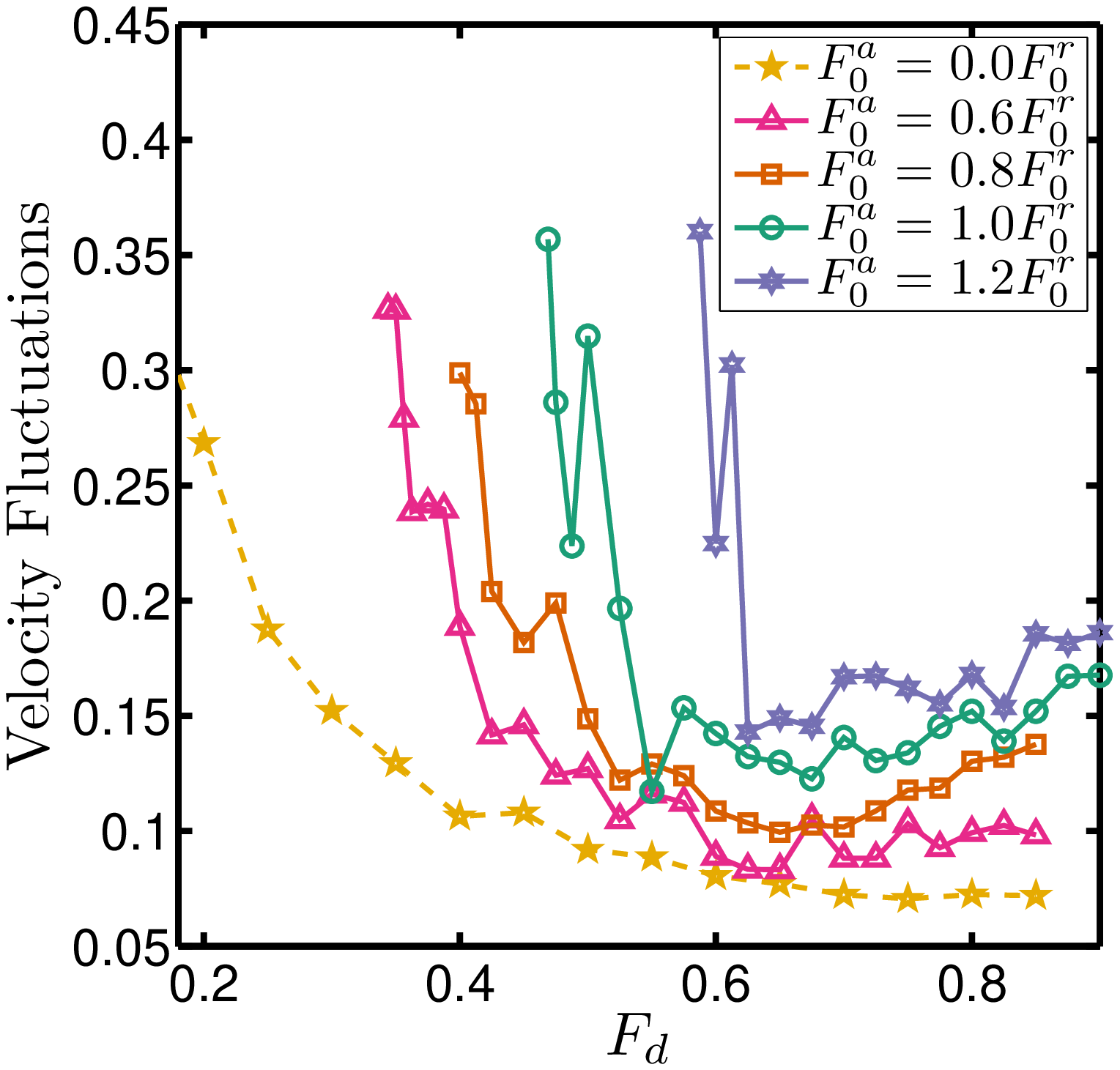}
\caption{The fluctuations in velocity, calculated as the standard deviation scaled by the mean, as a function of $F_d$. The large drop in velocity fluctuations is indicative of the transition from stick-slip to steady flow.}
\label{vel_fluct}
\end{figure}

The time-averaged centerline velocity at $x = 0.5$ as a function of driving force is shown in Figure \ref{meanvel_vs_attr}, plotted for different values of the attraction constant. For each value of attraction, we see that there is a driving force below which there is no steady state flow. We refer to this threshold driving force as the unjamming force $F_d^u$. The inset in Figure \ref{unjam_vs_attr}  shows that $F_d^u$ varies linearly with the strength of the attraction. We have checked by repeated simulations that the value of $F_d^u$ remains the same to below $2\%$ with different random initial bubble distributions. This plot has a non-zero intercept, meaning that for zero attraction, which corresponds to the unmodified Bubble Model, there is also a finite value of driving force below which there is no flow. This is the yield force. We distinguish this from the unjamming force $F_d^u$ at non-zero attraction force. The latter, we find, is not a conventional yield force, because we see some transient flow below the unjamming force (above the yield point) before the system jams, whereas below the yield point there is no flow. Another interesting aspect of the unjamming point is that it depends on how the initial state is prepared. We have tested this as follows: we begin to drive an initially equilibrated state with a constant force $F_d<F_d^u$ until the system eventually reaches a jammed state.  We then steadily ramp up the driving force, driving the system for a fixed time interval for each force, until we reach a force where the configuration "unjams", or shows a resurgence of nonzero flux. It is seen that the driving force at which this occurs is higher than that prescribed by the inset of Figure \ref{unjam_vs_attr}. Thus there is dependence on initial conditions, in the sense that the unjamming force required for a random equilibrated state is lower than that for a driven initial state that has become jammed.

The main plot in Figure \ref{unjam_vs_attr} shows data collapse when the centerline velocity from Figure \ref{meanvel_vs_attr} is plotted as a function of the driving force, $F_d$, scaled by the unjamming force, $F_d^u$, for each attraction strength. The collapse is particularly clean for higher driving forces, where the flow is smoother, whereas at driving forces just above $F_d^u$, it works less well - we identify this as the stick-slip regime. Hence the mobility depends purely on driving force for higher driving forces, while the attractive potential plays a more important role closer to the unjamming point. 

The corresponding relative fluctuations about the mean centerline velocity (the standard deviation scaled by the mean), plotted in Figure \ref{vel_fluct}, show an abrupt drop at a driving force slightly higher than the unjamming force for each value of attraction. This sudden reduction is indicative of a transition from a stick-slip to a steady flow. This transition becomes sharper as the attraction strength increases. On the other hand, no abrupt transition is seen in the case of zero attraction. The relationship between fluctuations and driving force, which is analogous to the stress-strain behaviour of the flow, is a consequence of the non-Newtonian nature of foams: it is known \cite{katgert2013} that fluctuations increase as strain rate decreases. We find that an attractive potential evidently plays a role in this, and its contribution to the stress-strain behaviour is a subject of ongoing work. 

\begin{figure*}
\centering
	\begin{subfigure}{0.75\linewidth}
		\centering
		\includegraphics[scale = 0.40]{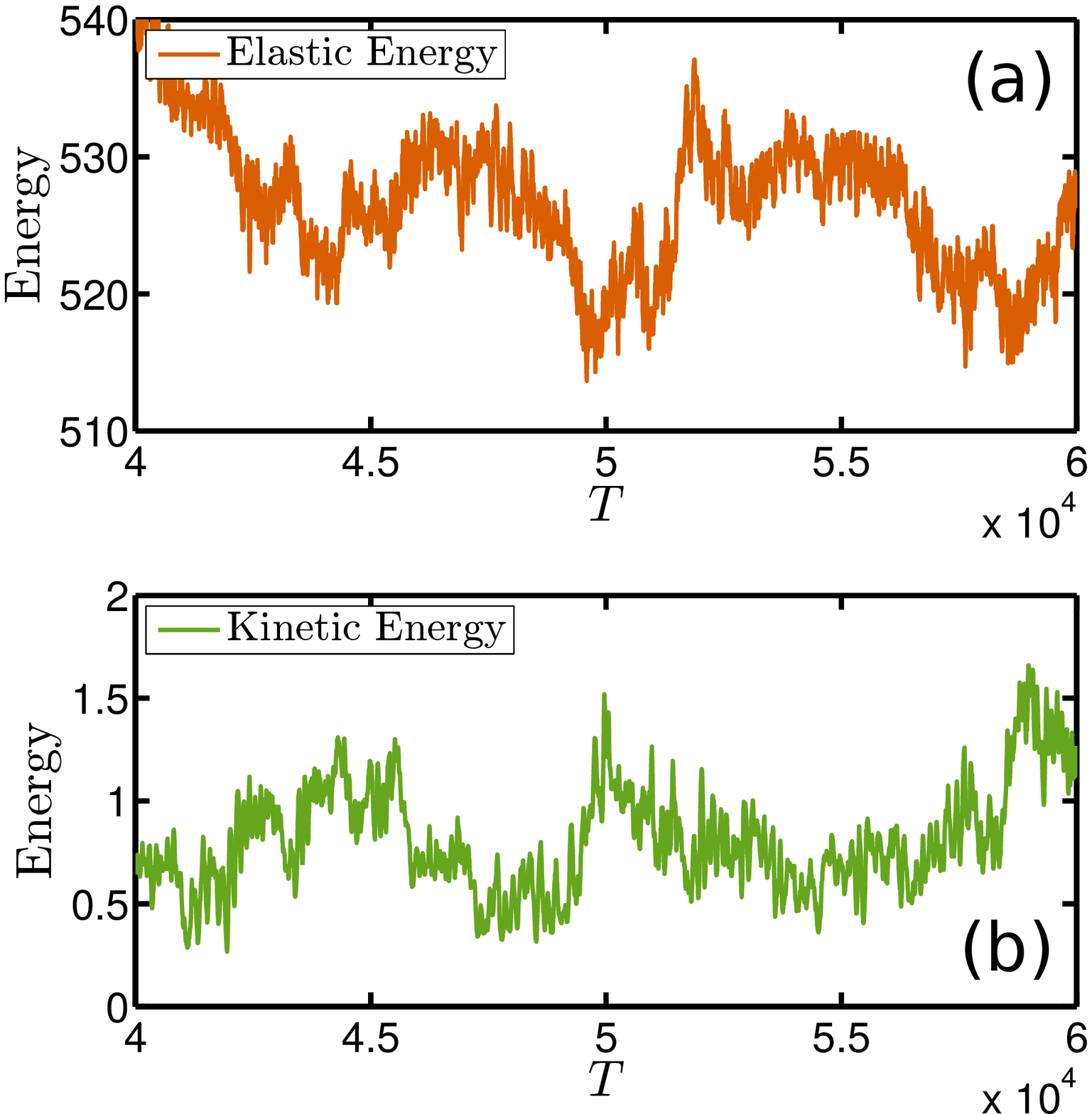}
		%\subcaption{$F_d = 0.50$}
		\label{kinetic_elastic_overlap_10}
	\end{subfigure}
	%\hspace{-1cm}
	\begin{subfigure}{0.45\linewidth}
		\centering
		\includegraphics[scale = 0.40]{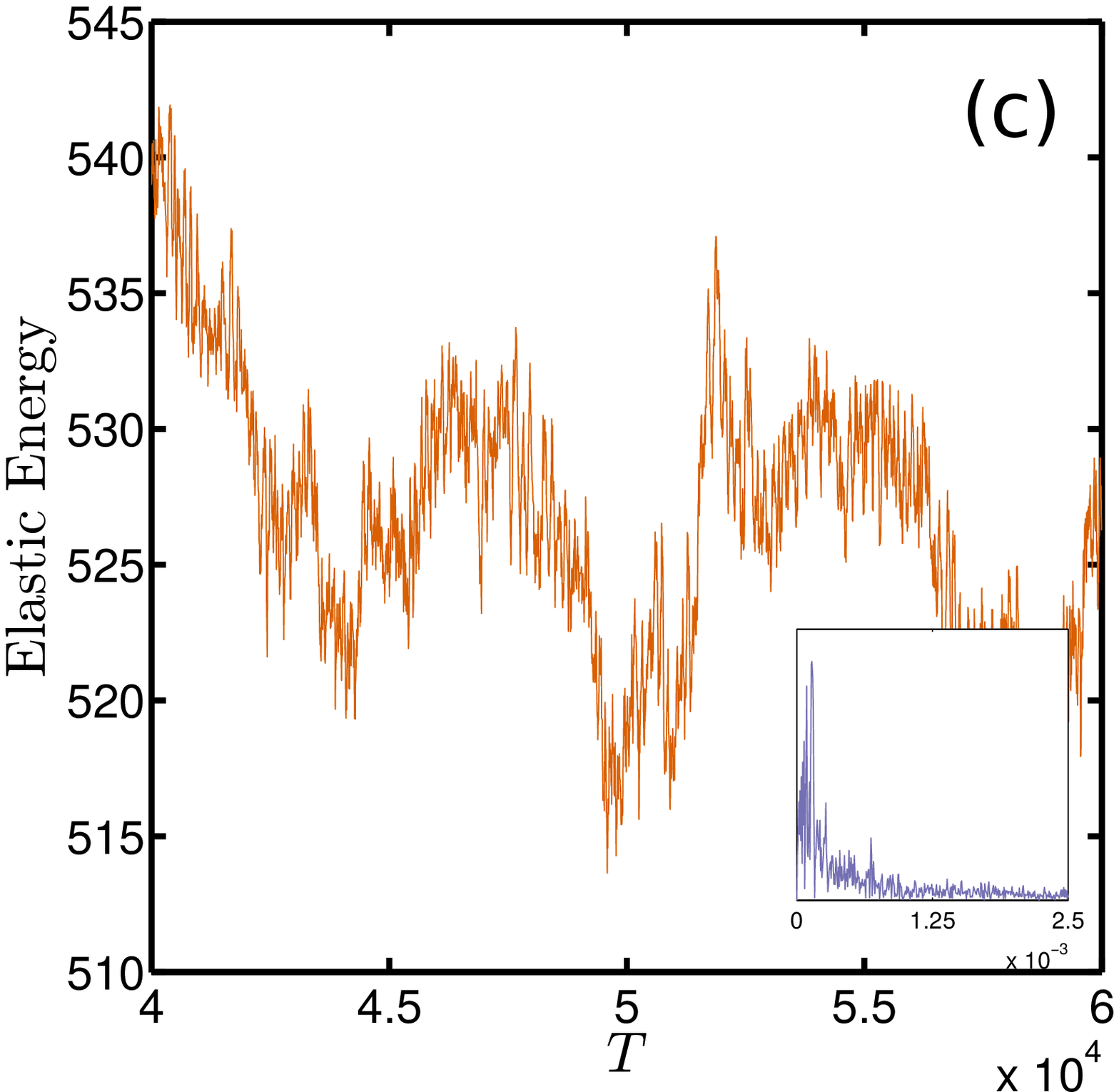}
		%\subcaption{$F_d = 0.50$}
		\label{elasticenergy_10_foutrans}
	\end{subfigure}
	%\hspace{3cm}
	\begin{subfigure}{0.45\linewidth}
		\centering
		\includegraphics[scale = 0.40]{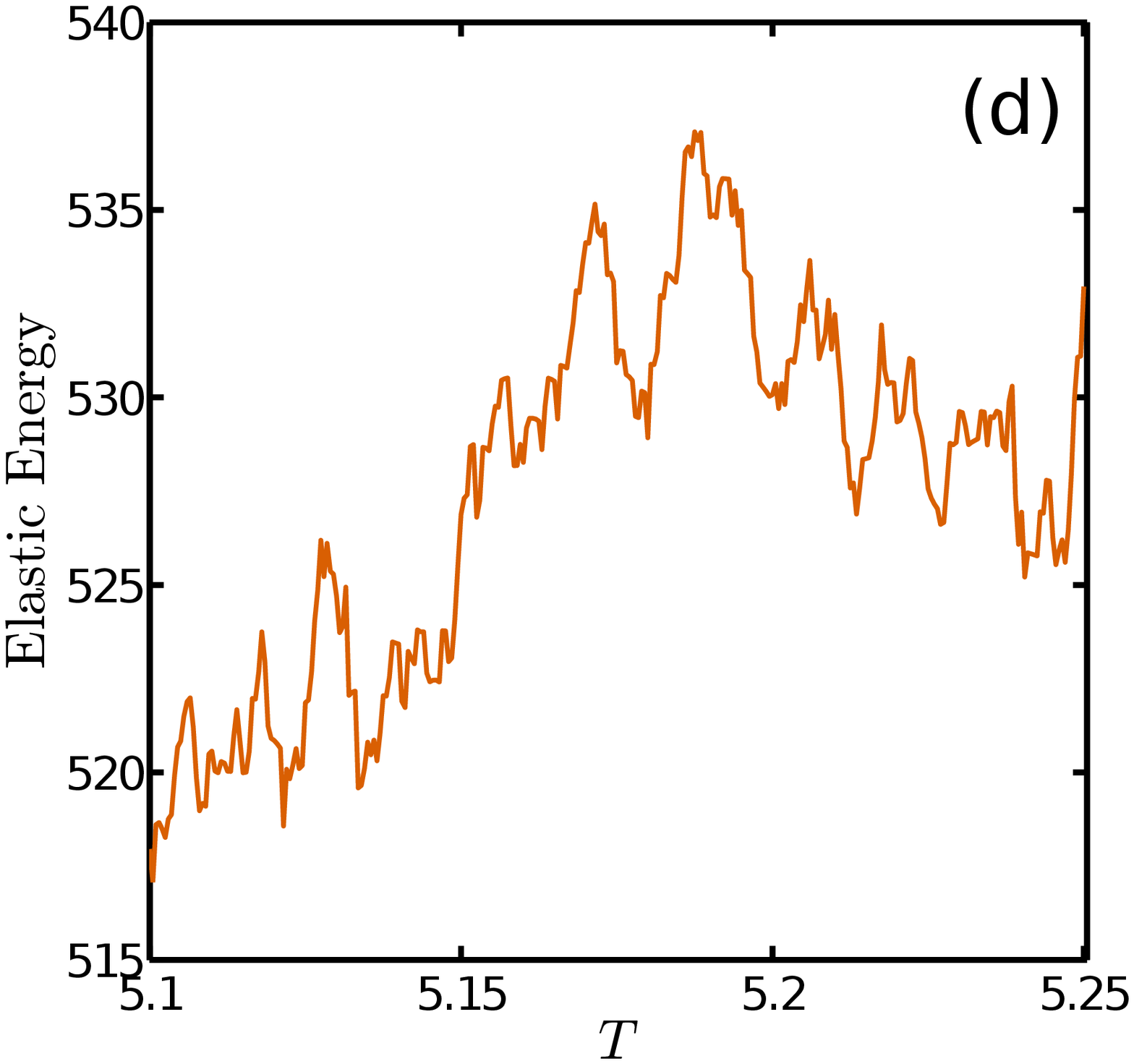}
		\label{elasticenergy_10_zoom}
	\end{subfigure}
	\label{elasticenergy_10}
	%\hspace{8cm}
	\begin{subfigure}{0.45\textwidth}
		\centering
		\includegraphics[scale = 0.40]{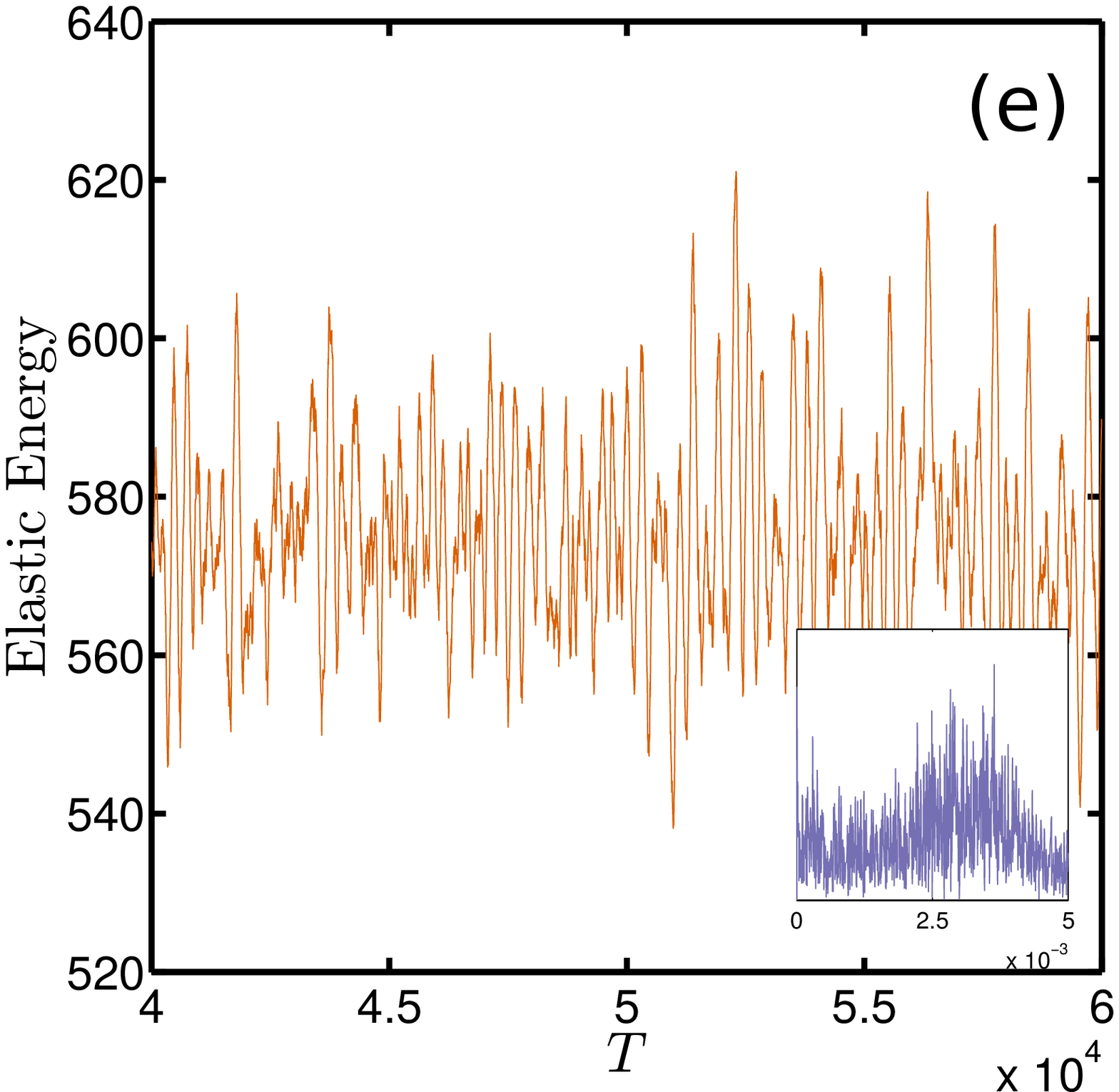}
		%\subcaption{$F_d = 0.90$}
		\label{elasticenergy_18_foutrans}
	\end{subfigure}
	%\hspace{3cm}
	\begin{subfigure}{0.45\textwidth}
		\centering
		\includegraphics[scale = 0.40]{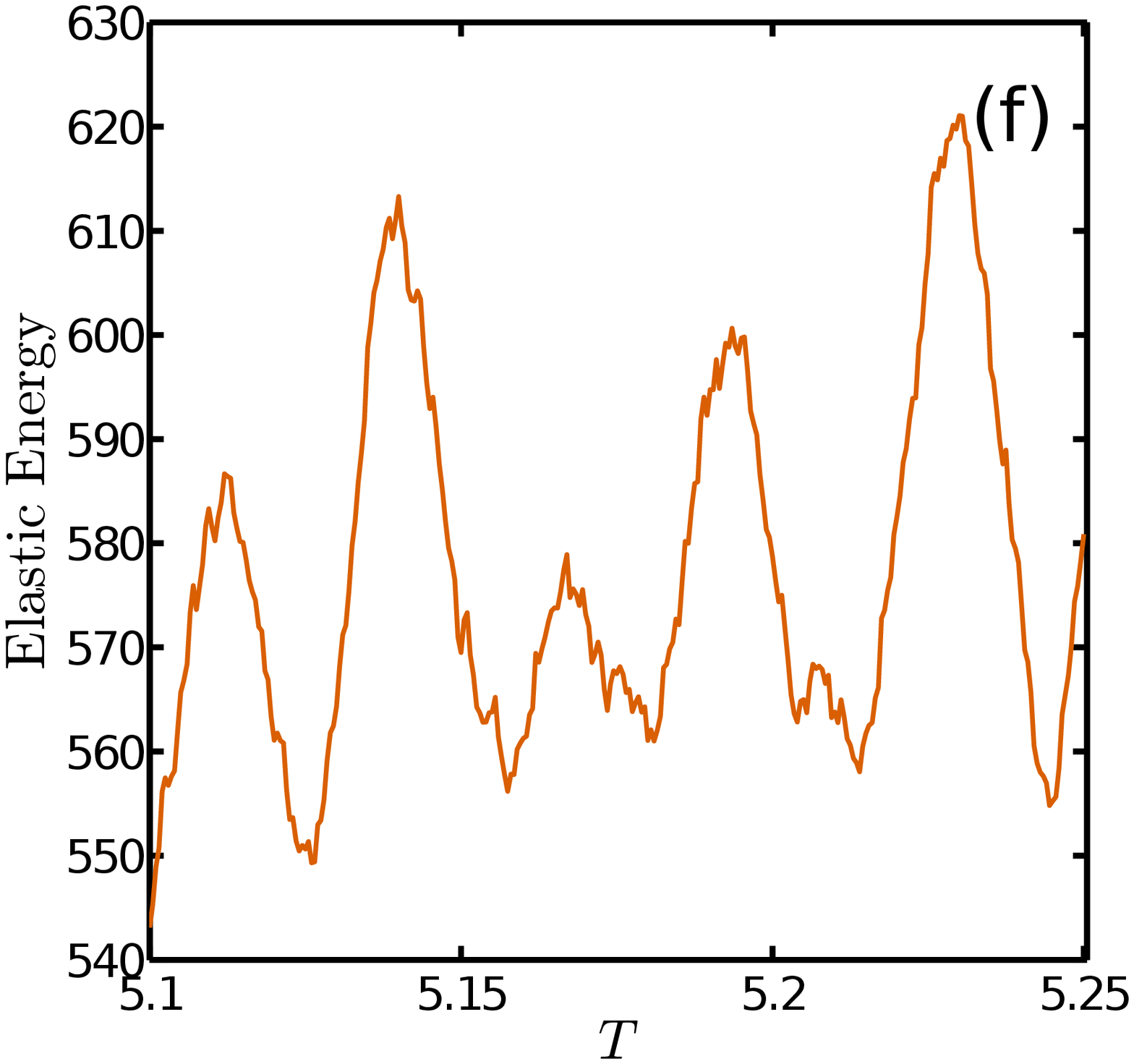}
		\label{elasticenergy_18_zoom}
	\end{subfigure}

\caption{ Fluctuations of elastic energy with time. Elastic energy in (a) is compared with kinetic energy in (b) to show their inverse correlation. Here $F_d = 0.5$; Elastic energy (c \& d) for $F_d = 0.5$; (e \& f) for $F_d = 0.9$. Figures (c) \& (e) show the behaviour over long time scales, and highlight the contrast between the slow loading and energy drops observed in stick slip flow and the frequent fluctuations in steady flow. The insets show Fourier transforms of these energy fluctuations. Figures (d) \& (f) show a zoomed-in picture of this behaviour at short time-scales.}
\label{energy_foutrans}
\end{figure*}

To better understand the effect of increasing driving force on the resulting flow, we  analyze the relative magnitudes of the stored elastic vs kinetic energy in the flow. The total elastic energy in the system, calculated as $0.5 k_r \delta^2_{ij}$ for each overlapping pair, and the total kinetic energy are shown as functions of time in Figures \ref{energy_foutrans}a and \ref{energy_foutrans}b respectively. It is evident that at a given time, the two are inversely correlated: kinetic energy increases when elastic energy drops. Thus fluctuations in elastic energy are sufficient to give a sense of the dynamics of the flow. These temporal fluctuations of elastic energy are shown in the lower half of the same figure, for stick-slip flow in (c) and (d) and steady flow in (e) and (f). Figures \ref{energy_foutrans}c, \ref{energy_foutrans}d show the elastic energy at low driving force, $F_d = 0.5$, the only difference between the two plots being that the one on the right, Figure \ref{energy_foutrans}d, has an expanded time axis. At low driving, large energy-drop events at long time scales dominate the flow. In this stick-slip regime, the driving force works to slowly build up the energy in the system, and the main flow is triggered by larger and less frequent energy release events. (The short time scale fluctuations correspond to the jitter of bubbles, where there is little collective movement or flow at the global level.)  In comparison, the flow at higher driving, $F_d = 0.9$ (Figures \ref{energy_foutrans}e, \ref{energy_foutrans}f), shows frequent energy drops of similar magnitude, as the expanded timescale on the right hand side plot confirms.

The insets in Figures \ref{energy_foutrans}c, \ref{energy_foutrans}e show the corresponding Fourier transforms of these energy fluctuations. As expected, the spectrum of energy fluctuations at low $F_d$,  Figure \ref{energy_foutrans}c inset, is peaked at low frequencies, which corresponds to the long timescale of energy-drop events. The steady flow case at higher driving force, Figure \ref{energy_foutrans}e inset, has a much noisier spectrum, with a broad peak at higher frequencies, indicating that the flow is dominated by faster energy drop events spread over a range of time scales.  

The timescale of the flow reponse is quite different in the stick-slip and steady flow regimes. The scale of fluctuations relative to the mean in these two cases, which correspond to the $F_0^a = 1$ curve in Figure \ref{vel_fluct}, are vastly different although the total number of bubbles in the domain remain the same. At the same time, a comparison of Figure \ref{energy_foutrans}c and \ref{energy_foutrans}e makes clear that the magnitude of the total elastic energy does not change significantly as the driving force increases. Hence only a fraction of the increased energy input into the system seems to contribute to elastic loading, while a majority of this energy is directed elsewhere. The steady flow is therefore characterized by largely collective and non-local dynamics, whereas the stick-slip regime is confined to local motions that do not seem to occur collectively. Simulation movies of the motion of bubbles, along with corresponding elastic energy fluctuations have been included as supplementary material, for the stick-slip as well as steady flow cases. 

\begin{figure}[h]
\centering
\begin{subfigure}{0.5\textwidth}
		\centering
		\includegraphics[scale = 0.4]{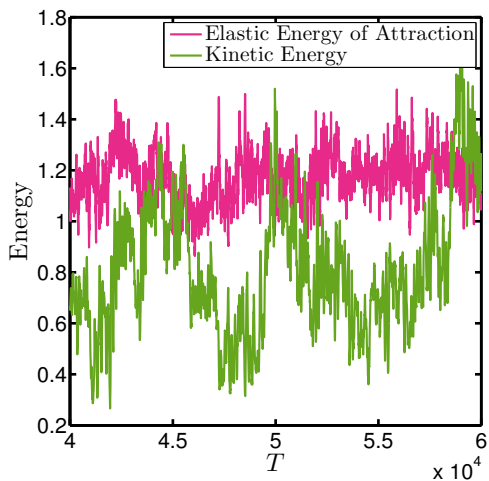}
		\captionsetup{justification=centering}
		\caption{$F_d = 0.5$. }
		\label{energycompare_10}
	\end{subfigure}
	\begin{subfigure}{0.5\textwidth}
		\centering
		\includegraphics[scale = 0.4]{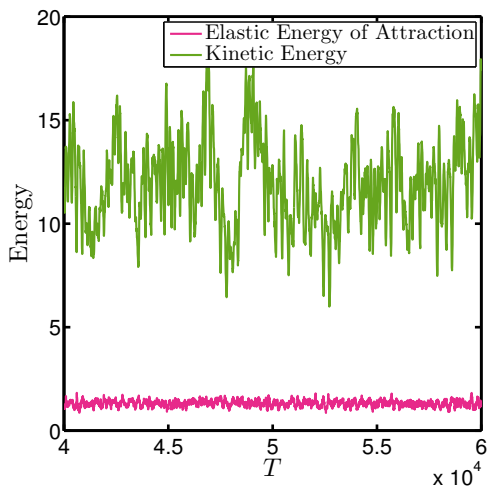}
		\captionsetup{justification=centering}
		\caption{$F_d = 0.9$.}
		\label{energycompare_18}
	\end{subfigure}
	
\caption{ A comparison between elastic energy of attraction and kinetic energy for (a) $F_d = 0.5$ and (b) $F_d = 0.9$. The elastic energy is similar in the two cases, but the kinetic energy is an order of magnitude larger for the higher driving in (b). }
\label{energycompare}
\end{figure}

The change in the nature of the dynamics is borne out by looking at how the distribution of driving energy changes with the strength of the driving force. Figure \ref{energycompare} shows a comparison of the stored attractive potential energy with the total kinetic energy at the same two driving forces used in Figure \ref{energy_foutrans}. We see that the total kinetic energy changes by an order of magnitude, while the attractive potential energy remains unchanged. At the lower driving forces, Figure \ref{energycompare}a, the contact networks formed due to the inter-bubble attractive potential are broken only when the kinetic energy and attractive potential energy are comparable in magnitude. At higher driving, Figure \ref{energycompare}b, the total kinetic energy is much larger than the attractive potential energy, and the contact networks are constantly being broken and re-formed. 

\begin{figure}[h]
\centering
\includegraphics[scale=0.4]{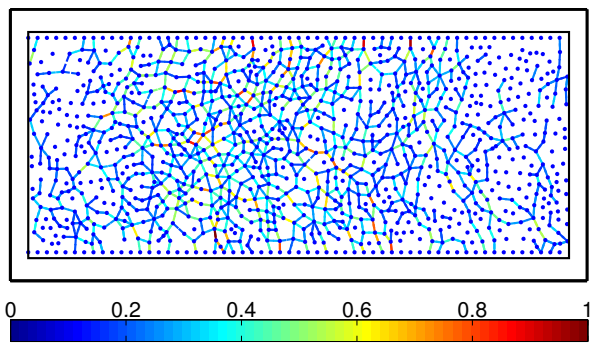}
\caption{A snapshot of the contact network in the system, shown at $F_d = 0.5$, where contact lines are plotted for overlaps, $\delta_{ij}$, exceeding $5\%$ of the sum of the radii. The colour corresponds to the degree of overlap, and the dots are bubble centres. }
\label{contact_network}
\end{figure}

\begin{figure}[h]
\centering
\includegraphics[scale=0.4]{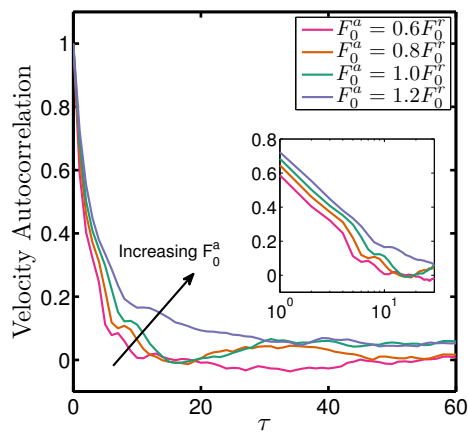}
\caption{Velocity autocorrelations at driving force $F_d = 0.8$ for different values of attraction strength. The decay time increases with attraction, highlighted in the inset by the linear-log plot. }
\label{autocorr_vs_attr}
\end{figure}

A comparison of the time evolution of the contact networks at low and high driving pressure shows that increased attraction between bubbles results in jamming behaviour. This evolution is shown as simulation movies for the low driving as well as high driving cases. For higher driving pressures, contact networks are frequently formed and broken, due to which the bubbles in the flow never find themselves trapped within a network. On the other hand, when the driving pressure is low, the contact networks formed are more persistent and bubbles flowing with lower velocities get trapped within these networks. A representative snapshot of the contact networks in a stick-slip flow is shown in Figure \ref{contact_network}. Some dense networks and trapped bubbles are seen here, but a clear qualitative discinction between the flow regimes is not evident in snapshots, and can only be seen in movies of the networks evolving with time (included as supplementary material). The slowing down and trapping of some bubbles within these networks leads to the formation of denser networks that persist for longer durations and subsequently cause more obstruction to the flow. This is supported by Figure \ref{autocorr_vs_attr} which shows a plot of the autocorrelation of bubble velocities for different values of the attraction strength: higher attractions are associated with longer decay times.

\begin{figure}
\centering
\includegraphics[scale=0.4]{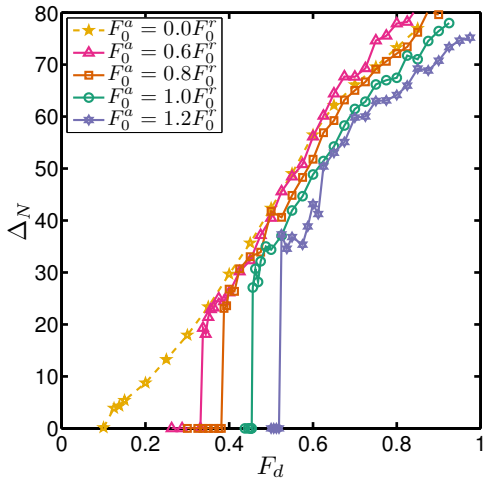}
\caption{The average number of contact changes, i.e., nearest-neighbour changes, per time step as a function of $F_d$, plotted for different values of attraction.}
\label{ngchange_vs_attr}
\end{figure}

\begin{figure}[h]
\centering
		\includegraphics[scale = 0.42]{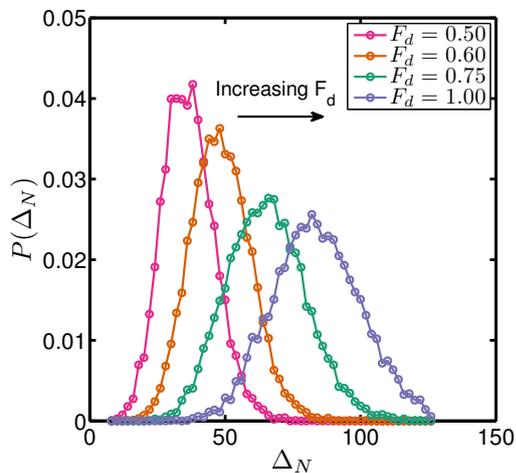}
		\caption{ Probability distribution for the number of neighbour changes per time step, shown at different values of $F_d$, for the attractive constant $F_0^a = 1.0 F_0^r$. The increasing width of the distributions signifies increasingly non-local behaviour. }
		\label{ngchange_dist}
\end{figure}

The average number of broken contacts per timestep, $\Delta_N$, is plotted as a function of the driving force in Figure \ref{ngchange_vs_attr}. For non-zero attraction, this rises abruptly from zero at the same threshold force at which flow is initiated, see Figure \ref{meanvel_vs_attr}, and then increases as driving force increases. For a given driving force $F_d$, there are fewer broken contacts on average as the attraction strength increases.  Figure \ref{ngchange_dist} shows the probability distribution of contact changes per timestep for different values of driving force for the case $F_0^a=F^r$. The standard deviation of this distribution increases with driving force, as is evident from the increasing width of the distributions. Plots of the standard deviation versus driving force, which are not shown here, overlap for different values of attraction, implying that the width of neighbour change distributions is independent of attraction. This trend in the number of broken contacts suggests an increasing non-local behaviour at higher driving forces, and suggests the contact forces are less restrictive as the driving force is increased. With higher driving force, as the dynamics seem to be confined by contact networks to a smaller extent, there is an increased interplay between various temporal and spatial scales of the flow. This non-local behaviour was also seen in the time series of energy fluctuations for the steady flow, and seems to be an important factor in the evolution from stick-slip to steady flow behaviour. 

\section{Conclusions}

We have investigated the effect of an attractive interaction between bubbles within the context of a model foam. This potential, ignored in most computational models of foams, is shown to introduce new flow regimes in channel flow. In particular, it leads to a regime of stick-slip behavior above the unjamming point. We have shown that this stick-slip flow is characterized by intermittent energy-release events. The scale of energy fluctuations suggest that these are local events, in contrast with the non-local behaviour observed in steady flow. It is possible that this difference is a consequence of the rigidity imparted to contact networks by the attractive potential. As we have shown, there is an interplay between kinetic energy and the elastic energy of attraction at low driving forces. This inability of the bubbles to break contacts at low driving forces could play a role in restricting the scale of rearrangements, hence forcing the dynamics to be confined locally. On the other hand, since contact networks are more easily overcome during steady flow, they do not play a restrictive role in the dynamics. Due to this, we see that the crossover from stick-slip to steady flow is accompanied by a reduced dependence of the mobility on the strength of attraction, and a drop in velocity fluctuations. 

The role of load-bearing contact force networks and their relationship to jamming have been studied in the context of shear-driven disordered systems such as dry granular media. The contact networks that seem to play a central role in jamming in our system are qualitatively different from those seen in granular systems - we see no signature of anisotropic force distributions such as force chains. An important consequence of the attractive potential is its ability to stabilize contact networks even in a simple geometry like a parallel channel. The rigidity imparted by the attractive potential is thus an important player in the jamming of pressure driven foams. The structure of these contact networks is the subject of continuing work.

This work can further be extended to the study of more complicated flow configurations, where the disjoining pressure has a very significant role to play even for steady flow. It has been seen \cite{langloisjrheol2014} that purely repulsive models do not work well for flows in geometries that have expanding sections - this includes modelling physically relevant systems such as porous media. Hence using computational models that include attractive potentials of this form to make these studies more realistic, and the extension of this work to converging-diverging channels is the subject of ongoing work. 

\subsection*{Acknowledgements}
We would like to thank Brian Tighe and Narayanan Menon for helpful discussions.

%%%REFERENCES%%%
%\bibliography{reference} 

%\bibliographystyle{plain} 

\end{document}